\documentclass[12pt,preprint]{aastex}
\newcommand{\Mbh}{$M_{\rm BH}$}
\newcommand{\ASCA}{{\it ASCA\/}}
\newcommand{\Chandra}{{\it Chandra\/}}
\newcommand{\ROSAT}{{\it ROSAT\/}}
\newcommand{\re}{$R_{\rm e}$}
\newcommand{\solarM}{$M_{\odot}$}
\newcommand{\solarLB}{$L_{\rm B,\odot}$}
\newcommand{\solarMtosolarLB}{$M_{\odot}/L_{\rm B,\odot}$}
\newcommand{\solarMtosolarLR}{$M_{\odot}/L_{\rm R,\odot}$}
\newcommand{\solarMtosolarLI}{$M_{\odot}/L_{\rm I,\odot}$}
\newcommand{\totM}{$M_{\rm tot}$}
\newcommand{\starMtoLB}{$M_{\rm star}/L_{\rm B}$}
\newcommand{\totMtoLB}{$M_{\rm tot}/L_{\rm B}$} 
\newcommand{\totMtoLR}{$M_{\rm tot}/L_{\rm R}$}

\begin{document}

\shorttitle{Mass Distributions in NGC 1407}
\shortauthors{Zhang et al.}

\title {Probing the Mass Distributions in NGC 1407 and Its Associated Group with the X-ray Imaging Spectroscopic and Optical Photometric and Line-strength Indices Data}

\author{Zhongli Zhang\altaffilmark{1,2},
        Haiguang Xu\altaffilmark{1,2},
        Yu Wang\altaffilmark{1},
        Tao An\altaffilmark{3},
        Yueheng Xu\altaffilmark{4}
        and Xiang-Ping Wu\altaffilmark{2}}

\altaffiltext{1}{Department of Physics, Shanghai Jiao Tong
University, 800 Dongchuan Road, Shanghai 200240, PRC; 
e-mail: zebrafish@sjtu.edu.cn, hgxu@sjtu.edu.cn, yuwen\_wang@sjtu.edu.cn}
\altaffiltext{2}{National Astronomical Observatories, Chinese
Academy of Sciences, 20A Datun Road, Beijing 100012, PRC; 
e-mail: wxp@bao.ac.cn}
\altaffiltext{3}{Shanghai Astronomical Observatory, Chinese Academy of Sciences, 80 Nandan Road, Shanghai 200030, PRC; 
e-mail: antao@shao.ac.cn}
\altaffiltext{4}{Department of Physics and Astronomy, University of Leicester, University Road, Leicester LE1 7RH, UK;
e-mail: yx12@star.le.ac.uk}

\begin{abstract}

We present a study of the mass distributions in the bright E0 galaxy NGC 1407 and its associated
group by analyzing the high quality \Chandra~and \ROSAT~X-ray spectroscopic data. In order to 
probe the stellar mass distribution we calculated the B-band mass-to-light ratio profile by 
comparing the observed line-strength indices and multi-color photometric data with different 
stellar synthesis model predictions. Based on the recent survey results we also have modeled the 
contribution from other group members to the stellar mass. We find that the gas is single-phase 
with a temperature of $\simeq0.7$ keV within 1\re~(1\re~= 9.0 kpc), which is typical for elliptical 
galaxies. Outside 1\re~the gas temperature increases quickly outwards to $>1$ keV, indicating its 
group origin. With the high spatial resolution of \Chandra~we reveal that the X-ray surface 
brightness profile shows a central excess in the innermost region, and on both the total mass 
and dark matter profiles there is a flattened feature at about $^{<}_{\sim}1$\re, which 
coincides with the gas temperature transition from the galactic level to the group level. We 
speculate that this may be a mark of the boundary between the galaxy and group halos, as has 
been seen in some other cluster/group-dominating galaxies. The total mass and dark matter 
distributions within $0.85$\re~are cuspy and can be approximated by power-law profiles 
with indices of $\simeq2$, which are marginally consistent with the generalized NFW profiles with
$\zeta=2$. The mass in outer regions can be well fitted by a single NFW profile, and the derived 
concentration parameter c ($18.6\pm1.5$) is larger than the 68\% upper limit for a halo at $z=0$ 
with the given $M_{\rm vir}$. We find that the NGC 1407 group has a baryon-dominated core, while 
the mass in the $>1$\re~is dominated by dark matter. At the virial radius $r_{200}=572\pm118$ kpc, 
the inferred mass and mass-to-light ratio are 
$M_{200}=2.20\pm0.42\times 10^{13}$ \solarM~and
$M_{\rm vir}/L_{\rm B}=311\pm60$ \solarMtosolarLB,
respectively, showing that the NGC 1407 group is an extremely dark system even comparable to many
clusters of galaxies. Since the obtained total mass is lower than those given in the earlier
galaxy kinematic works, we speculate that NGC 1400 is not a virialized member in the group's 
gravitational potential well.

\end{abstract}


\section{Introduction}
It has been well established that cold dark matter dominates the gravity on large scales 
in our universe. This is also true in clusters of galaxies for which observations via galaxy 
kinematics, X-ray imaging spectroscopy and lensing techniques have all given unassailable 
evidences that as much as about $80-90$\% of the gravitating mass is made of dark matter (see, e.g., 
Arnaud 2005 for a recent review). On galactic scales, kinematic studies of spiral galaxies 
via measurements of the rotation curves obtained with the radio HI and CO lines, as well as 
the optical H$_{\alpha}$ and [OIII] lines also have revealed the existence of a significant 
amount of unseen mass whose fraction rises steadily outwards (Sofue \& Rubin 2001). The 
mass-to-light ratios deduced from the observed constant rotation curves typically increase 
up to a few tens \solarMtosolarLB~within the last observed point, inferring a fraction 
of dark matter ranging from about 50\% in the Sa and Sb galaxies to about 90\% in the Sd 
and Sm galaxies.

Within the frame of the hierarchical models of galaxy formation and the theories of galaxy 
mergers, such a remarkable amount of dark mass is also expected in elliptical galaxies, especially 
in those luminous ellipticals who are more massive than their spiral counterparts. However, 
in elliptical galaxies the stars move in random, usually non-circular orbits, and there is a 
lack of obvious rotation velocity tracers like bright HI regions, particularly at large projected 
radii ($> 1$\re, where \re~is the effective radius) where the dark matter is considered 
to be important. So technically it is hard to model the 3D radial gradients and anisotropies 
in the velocity tensor field. In addition, in a bright elliptical galaxy the diffuse gas emission 
is usually heavily contaminated by the emission from a large population of low-mass X-ray binaries 
(LMXBs). Instruments with high spatial resolution and large effective areas are needed to disentangle 
the two components from each other accurately in the X-ray study of the gravitational potential well.

Despite of the difficulties, however, some early stellar photometric and kinematic works 
(e.g., Saglia et al. 1992; Bertin et al. 1994; Carollo et al. 1995) have indicated that 
in some ellipticals there may be as much dark matter as the luminous component within $\sim$\re. 
The derived mass-to-light ratios range from a few to a few tens of \solarMtosolarLB~with 
a tendency that the dark matter overwhelms the luminous matter at large radii, which supports
the possible homology of the galaxy formation processes between the ellipticals and spirals
(Bertola et al. 1993). As the absorption line profile measurements in elliptical galaxies improve
in recent years, the uncertainties in the rotation and velocity dispersion analyses caused by 
the degeneracy between anisotropy and mass distribution have been mitigated. Gerhard et al. (2001) 
found that the circular velocity curves (CVCs) of luminous ellipticals are flat to within 10\% 
in regions from $^{>}_{\sim}0.2$\re~out to 1-2\re, and they argued that about $10-40\%$ of the 
mass is dark within 1\re. In general, these kinematic and dynamic results are backed by the lensing 
measurements (Kochanek 1995; Griffiths et al. 1996; Keeton et al. 1998) and the X-ray imaging and 
spectroscopic observations (e.g., Forman et al. 1985; Mushotzky et al. 1994; Buote \& Canizares 
1994, 1998; Buote et al. 2002; Loewenstein \& White 1999; Fukazawa et al. 2006; Humphrey et al. 2006b),
as well as by the most recent photometric and dynamic analyses of more than 2000 SDSS galaxies 
(Lintott et al. 2005). Moreover, the superb high spatial resolution of the \Chandra~X-ray 
Observatory also allowed Buote et al. (2002) to examine the position angle twist and ellipticity 
of the X-ray image of NGC 720 with high precision. The authors argued that there is a strong 
geometric evidence for dark matter, which is independent of the temperature profile and is hard 
to be explained by the modified gravity theories as an alternative to dark matter (e.g., Milgrom 1983).

So far, however, unlike for the spiral galaxies the observational constraints on the dark 
matter content in ellipticals are diverse from case to case and, in some cases, unambiguous or
even controversial. For example, in the samples of Saglia et al. (1992), Bertin et al. (1994) 
and Gerhard et al. (2001) the kinematic and dynamic evidence for dark matter is found rather 
weak in some galaxies, which include the bright cD galaxy NGC 1399. A constant mass-to-light 
ratio cannot be ruled out with implication of no or less dark matter. In 2001 Baes \& Dejonghe
pointed out that after correcting for the dust absorption and scattering the need of dark matter 
in a few \re~may be reduced or eliminated in the kinematic study. Recently, Romanowsky et al. (2003) 
reported their analysis of the planetary nebula (PN) kinematics in the E1 galaxy NGC 3379, where 
the random stellar velocities are found to decline with radius. The calculated $M/L_{\rm B}$ is 
consistent with a constant, which infers a low dark matter fraction ($\leq 0.32$) at 5 \re, 
obviously conflicting with the standard galaxy formation theories. In order to respond to this 
challenge, Dekel et al. (2005) carried out merger simulations of disk galaxies and argued that 
the observed low random stellar velocities may be explained away in terms of the radial orbits 
of the halo stars that have been tidally ejected from the inner regions during merger events. 
No matter what the final conclusion may be, these examples clearly show the uncertainties in 
determining the dark matter content in elliptical galaxies.

A correct estimation of the stellar mass is crucial for probing dark matter distribution 
in early-type galaxies. In previous works it is sometimes assumed that stellar mass-to-light 
ratio is spatially constant throughout the galaxy, and in few cases a certain ``typical value'' 
was adopted. However, with the well-calibrated I-band data for 25 E/S0 galaxies Cappellari et al. (2006) 
showed that the stellar mass-to-light ratio distributions in the early-type galaxies show 
a remarkable variation from source to source by a factor of about 5, with a root mean square 
deviation of about 35\% for the sample. This variation is likely to be beyond the uncertainties 
induced in the stellar population model, which primarily arise in the line-strength calibration 
and the model assumption on the IMF profile and so on. Moreover, in a galaxy there is possibly 
a measurable spatial variation of stellar mass-to-light ratio on the $\sim$\re~scales, as is 
implied by the measured optical line index gradients (Rampazzo et al. 2005) and optical B-R color 
gradients (e.g., Tamura \& Ohta 2003, 2004; Forte et al. 2005). A straightforward example is
the kinematic study of 21 round elliptical galaxies by Kronawitter et al. (2000), who found 
that the stellar mass-to-light ratio may vary with radius by a factor of up to $\simeq2$ 
in a few galaxies. Padmanabhan et al. (2004) also reported a $^{>}_{\sim}30\%$ variation 
of the stellar mass-to-light ratio in some cases based on their comprehensive study of SDSS 
data.

As a step to get more insight into the dark matter distribution in early-type galaxies, in
this paper we present a multi-band study of the mass distributions in the bright elliptical 
galaxy NGC 1407 (E0, $z$=0.0059) and its associated group by incorporating both the spatially 
resolved X-ray spectroscopic data and optical multi-color photometric and line-strength data. 
In order to cover as wide a spatial range as possible, meanwhile, without the loss of the high 
spatial resolution in the central region, we drew the X-ray data from both the \Chandra~and 
\ROSAT~archives to probe the total mass and gas mass distributions, and drew the optical data 
from the literatures to calculate the stellar mass profile with a forward method. NGC 1407
is a non-cD, dominating galaxy that sits at the bottom of the group's gravitational potential 
well. According to the latest survey of Trentham et al. (2006), the group consists of 
about 250 member galaxies, of which about 85\% are dwarfs. The projected density of the globular 
clusters (GCs) falls off with the same slope as the halo light density of the central galaxy, 
indicating that the system may be dark matter-dominated (Perrett et al. 1996). By utilizing 
the velocities of 35 galaxies, Trentham et al. (2006) showed that the group has a total mass 
of $7\times10^{13}$\solarM~and a R-band mass-to-light ratio of \totMtoLR~=340\solarMtosolarLR. 
Note that NGC 1400 (non-barred SA0), the second brightest member that has an unusually large 
peculiar velocity of $1072~\rm{km~s^{-1}}$, was included in the calculation. The total mass of 
Trentham et al. (2006) is consistent with that of Gould (1993), however, it is larger than t
hat given in Quintana et al. (1994) no matter whether NGC 1400 is included or not. 
Based on the analysis of surface brightness fluctuations of Tonry et al. (2001), we adopted 
a distance to NGC 1407 of $26.8^{+3.4}_{-3.0}$ Mpc, which 
is slightly larger than that obtained with
$H_{0}=70$ km s$^{-1}$ Mpc$^{-1}$,
$\Omega_{m} = 0.3$ and $\Omega_{\Lambda} = 0.7$. 
We used the solar abundances of Grevesse \& Sauval (1998), where the iron abundance relative 
to hydrogen is $3.16\times10^{-5}$ in number.

\section{Observations and Data Reduction}
NGC 1407 was observed with the ACIS instrument on board the \Chandra~X-ray observatory on August 26, 
2000 for a total exposure of 50.3 ks. The CCD was operated at a temperature of $-110~^{\circ}{\rm C}$ 
with a frame time of 3.2 s. The center of the galaxy was positioned on the ACIS-S3 chip, with an 
offset of about $0.63^{\prime}$ from the nominal pointing for the S3 chip, so that most of the X-ray 
emissions of the galaxy was covered by the S3 chip. Although the ACIS I2-3, S1-2 and S4 chips were 
all in operation, we focused only on the data drawn from the S3 chip in this work. After removing the 
contamination of all the visually resolved points sources, we examined both the S1 chip and the 
boundary regions of the S3 chip where the emission of the diffuse gas is relatively weak. No background 
flare is found during the live operating time, which would otherwise significantly increase the count 
rate. We used the CIAO 2.3 software and included only the events with \ASCA~grades 0, 
2, 3, 4, and 6, and removed bad pixels, bad columns and node boundaries. The resulted clean exposure 
time is 48.5 ks. In order to constrain on the imaging and spectral properties of the outer regions, 
we also have analyzed the \ROSAT~PSPC data of NGC 1407 that was acquired from an observation 
performed on August 16, 1995, which lasted for 21.3 ks. We followed the standard \ROSAT~PSPC data 
analysis procedures to process the data by using XSELECT v2.3 and FTOOLS v6.0.3. The obtained clean 
exposure is 17.7 ks.

\section{X-Ray Morphology and Point Sources}
\subsection{X-ray Morphology}
In Figure 1a we show the smoothed \Chandra~ACIS S3 image of NGC 1407 in 0.3--10 keV in 
logarithmic scale, which is drawn from a 
$5^{\prime}{\times}5^{\prime}$ ($38.5\times38.5$ kpc)
region that is centered at the X-ray peak (RA=03h40m11.77s Dec=-18d34m49s J2000). 
The image has been corrected for exposure, but not for background. We find that the position 
of the X-ray peak is consistent with the optical center of the galaxy 
(RA=03h40m11.9s Dec=-18d34m49s; 2MASS, Jarrett et al. 2003) 
within about $1.8^{\prime\prime}$. The strong diffuse X-ray emission covers the whole S3 
field of view, and is roughly symmetric within  1\re~
(1\re~$=1.17^{\prime}$ or 9.0 kpc; de Vaucouleurs et al. 1991).
In $1-2$ \re, the spatial distribution of diffuse X-ray emission is less symmetric and is 
relatively stronger in the east, south and west. We estimate that within $\simeq2$\re~about 
3/4 of the diffuse X-ray has an origin from the hot interstellar medium (ISM), while the rest 
is from the unresolved point sources, most of which are LMXBs associated with the galaxy 
(Zhang \& Xu, 2004). Many point or point-like sources can be resolved visually in the \Chandra~
image, but only about half of them can be detected at the confidence level of $3\sigma$ (\S3.2). 
No remarkable irregularities or substructures such as X-ray filaments and cavities can be 
identified on the image.

In Figure 1b, we show the \ROSAT~PSPC image whose field of view 
($24^{\prime}{\times}24^{\prime}$, or $184.6\times184.6$ kpc) 
is much larger than that of the \Chandra~image. The image has been smoothed and exposure-corrected, 
too, but has not been corrected for background, either. Besides NGC 1407, the X-ray emissions 
associated with the group member NGC 1400 are also clearly detected. Being the second brightest 
group member in both optics and X-rays, NGC 1400 has a projected distance of about $11.7^{\prime}$ 
(90 kpc) southwest to NGC 1407 and has a much more compact appearance in X-rays. At the north-east 
of NGC 1400 there is a relatively weak, diffuse X-ray substructure aligned in the NGC 1400-NGC 1407 
direction (marked with a cross on the image), which is not identified on either optical or radio 
images. It may be related to a dwarf galaxy in the 
1407 group (Trentham et al. 2006) or may be a background source. Considering the unusually high 
peculiar velocity of NGC 1400, it is still possible that the structure is formed due to the gas 
stripping by ram pressure. We also identified a weak X-ray source associate with the background SB0 
galaxy NGC 1402 ($z=0.014$; Strauss et al. 1992). On the \ROSAT~image there shows some point and 
extended sources that have a 0.2--2 keV count flux larger than $0.001~\rm{cts~s^{-1}}$. None of 
them can be identified with either a Galactic or extragalactic source based on the currently 
available literatures. Because the study of these sources is beyond the scope of this work, we 
excluded them in the following imaging and spectral analyses. X-ray emissions from other bright 
group members are not detected.

Both \Chandra~and \ROSAT~images reveal that NGC 1407 and its host group are relaxed and are 
rich in hot gas. A detailed study of the \ROSAT~X-ray surface brightness profiles 
indicates that the X-ray gas extends outwards to at least $12^{\prime}$ (10.3 \re; \S5.1). The 
gas emission is soft, as is shown in Figure 1c where the smoothed 2-dimensional hardness ratio (HR) 
distribution calculated with the \Chandra~data is plotted after removing all the resolved point sources. 
The hardness ratio is defined as HR=(H-S)/(H+S), where S and H are the exposure- and background-corrected 
counts in 0.3--2 keV and 2--10 keV, respectively. The emission is very soft (HR=$-0.94 \sim -0.50$) 
within 1\re, which corresponds to a thermalized plasma at about $0.6-1.0$ keV. Outside 2\re~the 
hardness ratio increases to higher values that is typical for group gas. At about $1.8^{\prime}$ 
(13.8 kpc) east of the galactic center there is a relatively cold (HR=-0.7) region.

In Figure 1d we show the 1.43 GHz radio map of a $24^{\prime}\times24^{\prime}$ region around NGC 1407, 
which is obtained from the National Radio Astronomical Observatory (NRAO) archive. We identified NGC 1407 
as an extended source that has a flux density of $61.0\pm1.5$ mJy/b. The emission of the galaxy is 
asymmetric and extends farther towards west. A weak radio component, whose flux density is $9.6\pm2.2$ mJy/b 
($\sim3\sigma$), is detected at about $6^{\prime}$ southwest of NGC 1407. Also there is a bridge-like feature 
linking between the weak component and NGC 1407. In the 1.43GHz image we failed to identify the second 
brightest group member NGC 1400. However, this galaxy can be identified in the 0.3 GHz and 8.5 GHz maps 
(not shown here) with flux densities of $64.0\pm20.0$ mJy/b and $8.9\pm0.4$ mJy/b, respectively.
The lack of apparent substructures in the radio band such as remarkable jets and lobes strengthens our 
conclusion that the X-ray gas in NGC 1407 is relaxed."

\subsection{X-Ray Point Sources}
We detected X-ray point sources on the \Chandra~S3 image in 0.3--10 keV by using the CIAO 
tool {\it celldetect\/} with a threshold of $3\sigma$, and then crosschecked the results in 
0.7--7 keV and 2--10 keV separately by using both a wavelet arithmetic and visual inspection. 
After removing insignificant and fake detections, we totally detected 41 point sources within 
the central $\simeq2$\re, of which up to about 4 are probably unrelated background 
sources according to a Monte-Carlo simulation test based on the deep {\it Chandra} observations 
of blank fields (Mushotzky et al. 2000). We find that none of the resolved sources is identified 
with an optical counterpart. The spatial distribution of the point sources is nearly symmetric, 
which shows a clear concentration towards the galactic center. The source number density follows 
the B- and R-band optical light distributions approximately, indicating that most of them are 
intrinsically associated with the galaxy. In our previous work (Zhang \& Xu 2004) we studied 
the X-ray colors of these sources and their contribution to the total X-ray luminosity of the 
galaxy. We found that within 2\re~in 0.3--10 keV the resolved point sources account 
for about 18\% of the total emission; if the unresolved sources are taken into account, about 39\% 
of the total emission in 0.3--10 keV can be ascribed to LMXBs. The combined spectrum of all resolved 
off-center X-ray point sources can be well fitted with a single power-law model 
($\Gamma=1.59\pm0.09$) 
when the absorption is fixed to the Galactic value ($5.42\times10^{20}$ cm$^{-2}$; 
Dickey \& Lockman, 1990).

The brightest point source (Src 1) is detected at the X-ray peak of NGC 1407 and has a total 
net count of $364.4\pm21.0$ cts, or an average count rate of 
$75.1\pm4.3\times10^{-4}~{\rm cts~s^{-1}}$
in 0.3--10 keV. Previous to the \Chandra~observation, there is no report about the detection 
of this source in X-rays. We have attempted to detect the possible temporal variabilities of 
this source on different time scales. In terms of the K-S test, the temporal variability of the
source is less significant, possibly because K-S tests are most sensitive around the median 
value of the independent variable. However, the visual inspection of the lightcurve suggests 
that it is variable on timescales of 2.5-3 hr during which its count rate changed significantly 
by about $30-60\%$ (Fig. 2), which is supported by the calculated variability parameter 
$S=(f_{\rm max}-f_{\rm min})/\sqrt{(\sigma_{f_{\rm max}}^{2}+\sigma_{f_{\rm min}}^{2})}=1.9$,
where $f_{\rm max}$ and $f_{\rm min}$ are the maximum and minimum fluxes, respectively, and
$\sigma_{f_{\rm max}}$ and $\sigma_{f_{\rm min}}$ are corresponding errors. We fitted the ACIS spectrum 
of the central source with an absorbed power-law model. To avoid the relatively poor calibration 
at low energies and instrumental background at high energies, we limited the fittings to 0.7--7 keV. 
The absorption was fixed to the Galactic value since allowing it to vary did not improve the fit. 
The fit is acceptable ($\chi^2_r$=10.2/11) and the best-fit photon index is $2.29^{+0.24}_{-0.23}$ 
(Table 1), which is steeper than that of the accumulative spectrum of the off-center sources. We 
also have attempted to fit the spectrum of Src 1 with an absorbed multi-color disk blackbody model. 
However, the fitting is rejected with large residuals in both the low energy and high energy ends 
($\chi^2_r$=27.8/11). We estimate that the 0.3--10 keV luminosity of Src 1 is
$4.07\pm0.23\times10^{39}~\rm{erg~s^{-1}}$. 
This value is higher than that of the central sources in NGC 4472 and NGC 4649 (Soldatenkov et al. 2003) 
by one order of magnitude. Thus, we speculate that the central source may be an embedded low 
luminosity AGN because of its central location, rapid temporal variations, strong X-ray emissions 
and the appearance of a weak jet-like radio structure associated with it. If the source is an AGN, 
by using the tight relation between the black hole mass and the galaxy central velocity dispersion
$M_{\rm BH}=1.30(\pm0.36)\times10^{8}M_{\odot}(\sigma/200~\rm{km~s^{-1}})^{4.72(\pm0.36)}$ 
as was given in Merritt \& Ferrarese (2001), and the central velocity dispersion of $\sigma=272$ km s$^{-1}$ 
that is taken from McElroy (1995), we estimate the mass of the black hole as 
$5.55^{+2.37}_{-1.96}\times10^{8}$ \solarM. 
Quite similar results can be obtained if the \Mbh-$\sigma$ relation derived alternatively by 
Wyithe (2006) for a local sample of 31 galaxies and the central velocity dispersion of Bernardi et al.  
(2002; 279 km s$^{-1}$) are adopted instead.

\section{Spectral Analyses}
\subsection{Background}
In the \Chandra~spectral analysis of the ACIS S3 data the usual way to obtain a background 
spectrum, which primarily consists of the cosmic, instrumental and non X-ray particle components, 
is to utilize the observation dataset directly by extracting the background spectrum from a 
local, uncontaminated region on the S3 chip that is located far away enough from the source. 
Alternatively we may draw the background spectrum from the \Chandra~blank field datasets in 
the same detector region where the source spectrum is extracted. To avoid 1) the field-to-field 
variations in the cosmic background, which is typically of the order of a few tens percent 
on arcmin scales and even more remarkable on larger scales, and 2) the time-dependent variations 
in the particle and instrumental backgrounds that may be significant in some observations, the 
use of former method is more preferred in practice whenever it is applicable.

NGC 1407 is dominating a nearby group that is rich in X-ray gas. The diffuse X-ray emission
of the group spans over a wide region that covers the whole \Chandra~field of view. As we 
will show in \S5.1, the diffuse soft emission of the gas extends outwards to about 10.3 \re~($12^{\prime}$). 
At the edge of ACIS S3 chip, the emission from the hot gas is estimated to account for about 
$10-20\%$ of the total flux in 0.7--7 keV. Thus we are precluded from extracting a clean 
background spectrum directly in the CCD boundary regions. For the spectral fittings of the 
inner regions this will not be a problem; the X-ray background therein is overwhelmed by the 
emission of the galaxy in the selected bandpass, so even the \Chandra~blank field spectra 
can be applied as a good approximation of the real background if its high energy end, where the 
instrumental effects dominate, is well normalized to the observation. However, for the outer 
regions the use of the blank field spectra may result in severe misestimates of the model 
parameters due to the spatial and temporal variations of the different background components 
that may have changed their relative strengths. In fact, when we examined the spectrum 
extracted from the source-free region on the BI chip S1, we found that there is an extra 
hard component as compared with the blank field spectrum. Such a hard excess is also seen 
in the spectrum extracted in the boundary region of the S3 chip in 3--10 keV, where the gas
emission is relatively weak. The extra hard component found on the S1 and S3 chips are quite 
similar in spectral energy distribution, which can both be described by a Power-law model with 
$\Gamma=1.0$. The nature of this extra hard component is unclear. Possibly it is a feature 
of the local cosmic background, or it is induced by particle events. Therefore, the use 
of the blank field spectra should be rejected in our analysis.

In order to construct the \Chandra~S3 background spectrum, we extracted the spectrum of a 
$7.5^{\prime} \times 0.4^{\prime}$ region at the boundary of the S3 chip and then subtracted 
the contaminating soft emission component contributed by the hot diffuse gas from it. The 
selected region is located at about $5^{\prime}$ away from the galactic center, which is well 
beyond the outermost annulus used to extract the source spectra. As will be shown in \S4.2  , 
the contribution of LMXBs is negligible. We modeled the gas emission with an 
absorbed APEC component, whose temperature ($kT$=1.30 keV) and abundance ($Z$=0.21 solar) are obtained 
by studying the \ROSAT~PSPC spectrum extracted from the same region. Since in the \ROSAT~fittings the 
normalization of the APEC component is not well constrained, we deduced it from the modeling of the PSPC 
surface brightness profile (\S5.1). We find that in the selected background region the obtained ACIS S3 
background accounts for $82^{+5}_{-2}\%$ of the total counts in 0.7--7 keV and has a spectral 
energy distribution consistent with that obtained for the source-free region on the ACIS S1 
chip. The calculated background level also agrees with that derived in the analysis of the 
X-ray surface brightness profile (\S5.1) and that used in Humphrey et al. (2006b). In 0.2--2 keV 
where the particle and instrument components are minor, the deduced background count rate 
is close the average X-ray background count rate from the ROSAT All-Sky Survey diffuse background 
maps, which is available at http://heasarc.gsfc.nasa.gov.

For the \ROSAT~PSPC spectral analysis, we extracted the background spectrum from a region about 
$13^{\prime}$ away from the galactic center, where the 0.2--2 keV surface brightness (\S5.1) is 
less than $3\sigma$ above the mean local PSPC background level 
($7.26\times10^{-6}$ cts cm$^{-2}$ s$^{-1}$ arcmin$^{-2}$).

\subsection{Projected and Deprojected Analysis}
We performed both the projected and deprojected spectral analyses of the \Chandra~and \ROSAT~
spectra by using a model that consists of an APEC and power-law component, both subjected to a 
common absorption. In the deprojected analysis, we used the XSPEC model ``projct'' to evaluate 
the influence of the outer spherical shells onto the inner ones. In the \Chandra~spectral 
fittings, we extracted the ACIS S3 spectra from 4 annular regions that are centered at the X-ray 
peak. We limited the energy range to 0.7--7 keV to minimize the effects of both the instrumental 
background at higher energies and the calibration uncertainties at lower energies. By applying 
the latest CALDB we also have corrected for charge transfer inefficiency (CTI) and continuous 
degradation in the ACIS quantum efficiency, which is especially severe at low energies. To 
compensate for the degradation of the ACIS energy resolution we included an additional 5\% 
systematic error.  We first restricted the fittings to 3.5--7 keV, where the gas emission 
is typically less than $5\%$ and thus can be ignored. We found that an absorbed power-law model 
can give an acceptable fit to the spectra, and the obtained absorptions and photon indices are
consistent with the Galactic value and photon index for the resolved off-center point sources 
($\Gamma=1.59$; \S3.2), respectively. Thus, to achieve a better statistics in the fittings of
the 0.7--7 keV spectra, we fixed the photon indices to 1.59 and adopted the normalizations of 
the power-law components as have been obtained in 3.5--7 keV. Again, we fixed the absorption 
to the Galactic value because if we allow it to vary, the fitting did not improve. The best-fits 
are shown in Table 1. We see that the deprojected analysis gives acceptable fits, while the 
projected fits are poor for the inner three annuli. The residual distributions indicate a 
second thermal component needed for these annuli, which is most probably caused by the 
projection effect.

We extracted the \ROSAT~PSPC spectra in 0.2--2 keV from 8 annular regions, with the 4 inner annuli 
chosen as the same as for the \Chandra~analysis and the outer 4 annuli spanning over $2.1-10.3$ \re. 
We find that the soft emission of the gas always dominates in the spectra. For the inner 4 regions, 
we estimate that the hard emission component, which is mostly from LMXBs, account for up to $\simeq20\%$ 
of the 0.2--2 keV flux. Technically it is difficult to disentangle such a weak, hard component from 
the dominating soft emission component precisely in the limited bandpass of PSPC. Therefore when we 
fitted the \ROSAT~spectra of the inner 4 regions we always fixed the ratio of hard component to soft 
component referring to the \Chandra~results. For the outer 4 annuli, there is evidence that the 
relative contribution of the hard component is even lower. Forbes et al. (2006) reported that in NGC 1407 
the globular cluster (GC) density decreases exponentially outside 1.4 \re, with only about 
$\sim20\%$ of the total $\simeq1300$ GCs distributed in $r>2.1$\re. Since in the non-cD early-type 
galaxies, typically a few tens percent of the LMXBs are found in GCs (Xu et al. 2005 and references 
therein), the low GC density in the outer regions suggests a significant lack of LMXBs in the same 
regions. Therefore we fitted the \ROSAT~spectra of the outer 4 regions only with an absorbed APEC 
component. Since the abundances of the outermost three annuli are poorly determined, we tied them 
together during both the projected and deprojected fittings. For the same reason we tied the temperatures 
of the fifth and the sixth annuli together for the deprojected fitting. The fittings are marginally acceptable or poor for the 
projected, and are acceptable for the deprojected analyses (Table 1).

\subsection{3D Temperature and Abundance Profiles}
In Figure 3, we illustrate the three-dimentional gas temperature and abundance distributions
based on our deprojected spectral analysis. Within 2.1 \re~where both the \Chandra~
and \ROSAT~data are available, the gas temperature and abundances obtained with the \Chandra~
and \ROSAT~agree very well with each other within the errors, except that in $0.53-2.1$\re~
the \ROSAT~temperature is lower. In the analyses that follows, we adopt the \Chandra~best-fits 
for the inner 4 annuli and the \ROSAT~best-fits for outer ones. We noticed that within 1\re, 
the gas temperature is below 1 keV but still higher than 0.6 keV, which is typical 
for bright early-type galaxies. The gas temperature increases steadily outwards, reaches a maximum 
value of 
$1.37\pm0.35$ keV in $2.1-5.1$\re, 
and then drops to about 1.1 keV in the outermost annulus. The high temperature in $^{>}_{\sim}1$\re~
suggests the group origin of the gas. We have attempted to model the temperature profile with 
an analytical form, and found that
\begin{equation}\label{eq:temp_model} 
kT=\left\{\begin{array}{ll}
c_0 +\frac{c_1(\frac{r}{r_{\rm c}})^{\eta}}{1 +(\frac{r}{r_{\rm c}})^{\eta}} & \mbox{$r<3.4$\re} \\
\\
c_0 +\frac{c_1(\frac{34-r}{r_{\rm c}})^{\eta}}{1 +(\frac{34-r}{r_{\rm c}})^{\eta}} & \mbox{$r>3.4$\re} 
\end{array}\right.
\end{equation}
(Pratt \& Arnaud 2003) can best describe the observed temperature profile. The parameters are
$c_0=0.66$ keV, 
$c_1 =0.71$ keV, 
$r_c = 0.94$ \re~and 
$\eta=3.78$ 
for $r<3.4$\re, and 
$c_0=0.74$ keV, 
$c_1 =0.70$ keV, 
$r_c = 25.5$\re~and 
$\eta=11.22$ 
for the outer regions.
The abundance at the center of the galaxy is about 0.56 solar. It increases to about 1.40 solar 
at $\sim0.85$\re, and then decreases rapidly to 0.21 solar in $^{>}_{\sim}3.4$\re. We 
approximate the abundance distribution in the analytical form as 
\begin{equation}\label{eq:abun_model} 
Z = c_0 +c_1 \times {\rm exp}(\frac{r-r_{0}}{\sigma})^{-2},
\end{equation}
where $r_{0}=0.7$\re. Within 0.7 \re, we obtain 
$c_0=0.28$ solar, 
$c_1 =1.06$ solar and 
$\sigma =0.58$. 
In outer regions, the parameters are
$c_0=0.21$ solar, 
$c_1=1.13$ solar and 
$\sigma=1.12$. 
In general, our results agree with those of Matsushita (2001), Fukazawa et al. (2006) 
and Humphrey et al. (2006a,b) very well, although different annular-binnings were used in these works.

\section{Mass Distributions}
\subsection{X-ray Surface Brightness Profiles and Gas Distribution}
In Figure 4 we show the X-ray surface brightness profiles $S(r)$ of our target obtained with 
\Chandra~ACIS in 0.7--7 keV and \ROSAT~PSPC in 0.2--2 keV, which have been corrected for 
exposure but not for background. In order to improve the fitting statistics and to model the 
gas distribution over as wide a region as possible, we fitted the two profiles jointly. By 
assuming hydrodynamic equilibrium, spherical symmetry, and that the gas is ideal and thermalized, 
we integrated the gas emission along the line of sight by using
\begin{equation}
S(r) = S_{0} \int^{\infty}_{r} \Lambda (T,Z) n_{g}^{2}(R) \frac{RdR}{\sqrt{R^{2}-r^{2}}}+S_{\rm bkg},
\end{equation}
where $R$ is the 3-dimensional radius, $\Lambda(T,Z)$ is the cooling function calculated by 
taking into account both the temperature and metallicity gradients as are given by the best-fit 
deprojected spectral model in \S4.3, $n_{g}(R)$ is the gas density, and $S_{\rm bkg}$ is the 
background. For $n_{g}(R)$ we have attempted to apply either the $\beta$ model, which takes 
the form of
\begin{equation}
n_{g}(R) = n_{g,0} \left[ 1 + (R/R_{c})^{2} \right] ^{-3\beta/2},
\end{equation} 
or the two-$\beta$ model 
\begin{equation}
n_{g}(R) = \left\{
           n_{g,1}^{2} \left[ 1 + (R/R_{c1})^{2} \right] ^{-3\beta_{1}}
         + n_{g,2}^{2} \left[ 1 + (R/R_{c2})^{2} \right] ^{-3\beta_{2}}
           \right\} ^{1/2},
\end{equation}
where $R_{c}$ is the core radius, $\beta$ is the slope, and suffixes 1 and 2 are used to represent
the two emission components. For inner regions, we have removed the contributions from both resolved 
and unresolved point sources based on our \Chandra~spectral analysis (\S4.2). For the $>2.1$\re~ 
regions where the \Chandra~spectral information is not available, we estimate that the hard X-rays 
from the point sources accounts for up to about 5\% of the total flux in the \ROSAT~PSPC spectra, 
so we simply ignored this component. We also eliminated the faint X-ray substructure located between 
NGC 1407 and NGC 1400 (\S3.1).

We find that the fit with the $\beta$ distribution should be rejected on the 90\% confidence 
level ($\chi^2_r$=167.9/60), which shows significant residuals in the central $0.2$\re~for the \Chandra~
Surface brightness profile that infers the existence of a central excess (Figure 4a). The two-$\beta$ model, 
on the other hand, gives acceptable fit to the data ($\chi^2_r$=68.9/56; Fig. 4) with $R_{c1}=0.12\pm0.01$\re~
and $\beta_{1}=0.70\pm0.01$ for one component, and $R_{c2}=0.83\pm0.01$\re~and 
$\beta_{2}=0.45\pm0.01$ for another. Using the best-fit gas distribution we find that the gas density 
at the galactic center and at 1\re~are $\sim0.1~\rm{cm^{-3}}$ and $\sim0.002$ cm$^{-3}$, respectively. 
The gas density decreases to $\simeq10^{-4}$ cm$^{-3}$ at about $10.3$\re~where the count 
rate is about $3\sigma$ above the mean background level. The best-fit \Chandra~background level 
is $1.0\times10^{-5}$ cts s$^{-1}$ cm$^{-2}$ arcmin$^{-2}$ and is in excellent agreement with 
what we have adopted in the spectral analysis (\S4). We also notice that even if the \ROSAT~data 
of the inner regions are excluded from the fittings, the best-fit parameters do not change 
significantly at the 90\% confidence level. This is not surprising because central excess 
emission is essentially restricted in a region whose size is similar to the \ROSAT~PSPC PSF.

\subsection{Stellar Mass}
\noindent{\it NGC 1407}\\
\indent We first employed the PEGASA-HR code (Fioc \& Rocca-Volmerange 1997; Le Borgne et al. 2004) 
to deduce the stellar mass-to-light ratio \starMtoLB~in NGC 1407. By assuming the solar abundance 
ratios at the inferred age and metallicity, and utilizing an empirical stellar library that 
includes both the thermally pulsating asymptotic giant branch stars (AGB) and post-AGB phases, 
the code has been designed to degenerate the effects of age and metallicity by creating high 
resolution synthetic spectral energy distributions (R=10 000) that can be compared with the 
observed line-strength indices and photometric data. We quoted the B- and R-band photometric 
data of NGC 1407 from de Carvalho et al. (1991), both of which exhibit similar profiles to those 
provided earlier by Lauer (1985) and Franx et al. (1989), and drew the K-band photometric data from 
Jarrett et al. (2003). To calibrate the data, we normalized the B-, R- and K-band magnitudes 
within 1\re~to 
$\mu_B=21.79$ mag arcsec$^{-2}$, 
$\mu_R=20.24$ mag arcsec$^{-2}$ and 
$\mu_K=19.09$ mag arcsec$^{-2}$, 
respectively (Prugniel \& Simien 1996), and then corrected the data for 
Galactic extinctions (Schlegel et al. 1998). Like in many other elliptical galaxies (e.g., Burkert 1993), 
the B- and R- band surface brightness profile $S(r)$ in 0.1--0.7\re~and the overall K-band 
$S(r)$ can be described with the $1/4$ law 
\begin{equation}
I(R) = 8.0 S_{\rm e} 10^{-3.331 [(R/R_{\rm e})^{1/4}-1]}, 
\end{equation}    
where $S_{\rm e}=6.34\times10^{-11}~\rm{erg~s^{-1}~cm^{-2}~arcmin^{-2}}$ (B-band),  
$1.13\times10^{-10}~\rm{erg~s^{-1}~cm^{-2}~arcmin^{-2}}$ (R-band) and
$3.12\times10^{-10}~\rm{erg~s^{-1}~cm^{-2}~arcmin^{-2}}$ (K-band).
In the innermost region, however, both the B- and R-band $S(r)$ shows 
a clear excess beyond the $1/4$ law, which can be characterized by a ``Nuker'' model (Faber et al. 1997) 
\begin{equation}
I(r) = 2^{(b-c)/a} I(r_{\rm br})(r_{\rm br}/r)^{c}[1+(r/r_{\rm br})^{a}]^{(c-b)/a},
\end{equation}
where $a$ parameterizes the sharpness of the break, 
$b$ is the asymptotic outer slope, 
$c$ is the asymptotic logarithmic slope inside $r_{\rm br}$, 
and $r_{\rm br}$ is the break radius. We find that in the B-band 
$a=2.20$, $b=1.67$, $c=0.05$ and $r_{\rm br}=0.05$\re,
and in the R-band
$a=2.20$, $b=1.72$, $c=0.05$ and $r_{\rm br}=0.05$\re.
By modeling the observed surface brightness profile we calculated that the B-band luminosity 
within 1\re~is $3.5\times10^{10}$ \solarLB. 
We adopted the measured line indices in four luminosity weighted annuli ($0-1/16$\re, $1/16-1/8$\re, 
$1/8-1/4$\re~ and $1/4-1/2$\re), which include the emission- and velocity-corrected H$\beta$, 
Mgb, Fe5015, Fe5270 and Fe5335 indices (Table 2). Using the method described in \S4 of Rampazzo et al. (2005), 
the indices have been transformed into the Lick-IDS system, which is originally designed to 
describe the spectral absorption features by including a set of atomic and molecular bands, 
each consisting of a central bandpass and two pseudocontinuum bands flanked to the red and 
blue sides (more details about the Lick-IDS system is described in Trager et al. (1998)).  
The H$\beta$ line index was directly corrected by comparing with the H$\alpha$ lines other 
than by comparing with the [O III] lines, since the H$\beta$/[O III] correlation has a relatively 
larger dispersion that varies between 0.33 and 1.25 (Trager et al. 2000; Denicol\'{o} et al. 2005).

In the calculations we assumed that 1) the stellar population is coeval, i.e., the line-strength 
indices and color gradients are mostly due to the metallicity gradient (Michard 2005), 2) the star 
formation rate can be approximated by an exponentially decreasing form 
SFR(t)$\propto e^{-t/p}$, 
where $p$ is the decay time in Myr, and 3) the initial mass function (IMF) takes the form of 
either the Salpeter profile (Salpeter 1955) or the Kroupa profile (Kroupa 2001) in the mass range 
$0.1-120$\solarM. After considering the effects of dust, galactic wind, as well as the nebular 
emission, we compared the model-predicted line-strength indices, B-R and B-K colors with the 
observed data (Table 2). We found that magnesium is always underestimated by the model, which confirms the 
results of Humphrey \& Buote (2006a), Howell (2005) and Thomas et al. (2005) that NGC 1407 is an 
$\alpha$-enriched system. In order to reconcile the assumption of the solar abundance ratios 
in the PEGASE code with the $\alpha$ overabundance, we alternatively employed a combinative 
index MgFe as a tracer of total stellar metallicity, which is defined either as  
${\rm [MgFe]_{T}}=\sqrt{{\rm Mgb}(0.72\times{\rm Fe}5270+0.28\times{\rm Fe}5335)}$ (Thomas et al. 2003)
or as
${\rm [MgFe]_{C}}=(0.45Mgb+Fe5015)/2$ (Cappellari et al. 2006).
A recent study with excellent dynamical models and high quality spectra from the SAURON panoramic 
integral-field spectrograph by Cappellari et al. (2006) shows that the \starMtoLB~calculated 
with the so-defined MgFe index is weakly affected by non-solar $\alpha$/Fe. This is also shown
in Table 3 where the deduced age and metallicity of the stellar population of NGC 1407 in all 
annuli are insensitive to the choice of IMF and MgFe definations. Within the errors the stellar 
age and metallicity deduced with different combinations of IMF and MgFe agree well with each 
other. In all cases the derived stellar age is close to 10 Gyr, which is consistent with some 
previous works ($7.4\pm1.8$ Gyr, Thomas et al. 2005; $9.5\pm2.2$ Gyr, Howell 2005), and the 
metallicity always reaches its maximum value in $1/8-1/4$\re~and then decreases to a lower value 
in $1/4-1/2$\re. The \starMtoLB~deduced with the Slapeter IMF, on the other hand, is apparently 
larger than those deduced with the Kroupa IMF (Table 3) at the 90\% confidence level (see \S6.1
for a detailed discussion about the errors). We notice that the model predictions with the 
Salpeter IMF fit the observed data significantly worse than those with the Kroupa IMF. This 
may be caused by the fact that the Salpeter IMF is too rich in low-mass stars (e.g., 
Weiner et al. 2001). If a diet Salpeter IMF (Bell et al. 2003) is adopted instead, the 
\starMtoLB~values may be reduced by $\sim30\%$, which is in excellent agreement with the 
values deduced with the Kroupa IMF.

For the time being the line index measurements are not available for the regions outside 0.5\re.
Considering that both the 2MASS H-K and J-K colors keep approximately invariant within the 
error in 0.5-2\re~(Jarrett et al. 2003), which infers that the stellar mass-to-light ratio is 
roughly constant in the same region, we fixed the stellar mass-to-light ratio in $r>0.5$\re~
to 4.45 \solarMtosolarLB~as is derived in 0.25-0.5\re.

\noindent{\it Other group members}\\
\indent More than 250 galaxies have been identified in the NGC 1407 group, of which about 85\% 
are dwarfs. By referring to the data in Trentham et al. (2006) we find that the 2-dimensional galaxy 
number density roughly follows a power-law profile with an index of $\sim-2/3$. Utilizing the
mean B-band stellar mass-to-light ratio as have been obtained for NGC 1407, we modeled the 
stellar mass distribution in $<10.3$\re. We find that, excluding NGC 1407, all other member 
galaxies contribute about 25\% of the total stellar mass, or $\simeq$1.7\% of the total mass 
within 10.3 \re.

\subsection{Total Gravitating Mass}
Assuming that the gas is ideal and the group is in the hydrostatic equilibrium state, 
we calculate the gravitating mass distribution by using 
\begin{equation}
M_{\rm tot}(R) = -\frac{kTR}{G\mu m_{p}}\left(\frac{d\rm ln \it n_{g}}{d\rm ln \it R}
+\frac{d\rm ln \it T}{d\rm ln \it R}\right),
 \end{equation}
where $M(R)$ is the total mass within the radius $R$, $n_{g}$ is the gas density, G 
is the gravitational constant, $m_{p}$ is the proton mass and $\mu$ is the mean molecular weight. 
In the calculation we adopt the best-fit gas density profile obtained in the joint
\Chandra~and \ROSAT~fittings of the X-ray surface brightness profiles (\S5.1, Fig. 4) and the 
best-fit gas temperature profile (\S4.3) obtained in the spectral modelings (\S4.2). The resulting 
gravitating mass distribution is shown in Figure 5a, along with the 90\% errors determined by 
performing Monte-Carlo simulations that account for the ranges of temperature and gas density 
allowed by the data. The derived total mass of NGC 1407 is 
$2.75^{+0.47}_{-0.56}\times10^{11}$ \solarM~at 1\re~
and 
$6.6^{+1.2}_{-4.1}\times10^{12}$ \solarM~at 10.3 \re, the gas halo boundary. 
We noticed that at $R \approx 0.85$ \re~there is a clear change in the slope of $M(R)$, exhibiting 
a shoulder-like structure that has been observed in galaxy clusters  (e.g., Abell 1795, 
Xu et al. 1998 and Ettori et al. 2002) and other groups (e.g., the NGC 1600 group, Sivakoff et al. 2004). 
In Figure 5b we show the mass-to-light ratio \totMtoLB~profile. We find that at $^{<}_{\sim}0.85$\re~the 
lower and upper limits of the ratio are $\simeq5$ and $\simeq10$\solarMtosolarLB, respectively. 
At the boundary of optical envelope ($\simeq5.1$\re) 
\totMtoLB~is about 50\solarMtosolarLB, indicating the existence of a large amount of dark 
matter. At the gas halo boundary \totMtoLB~increases to over 90\solarMtosolarLB.

Due to the appearance of the baryonic component that dominates in the inner region, the total 
gravitating mass profiles of early-type galaxies are usually inconsistent with the dark halo 
profiles given in Navarro et al. (1997; the NFW profile) and Moore et al. (1999; the generalized 
NFW profile), which can be expressed as
\begin{eqnarray}
\rho(R)&  = & \frac{\rho_c(z)\delta_c}{(R/R_s)^{\zeta}(1+R/R_s)^{3-\zeta}}, 
\label{eqn.generalized_nfw}
\end{eqnarray}
where 
$\zeta<2$ ($\zeta=1$ for the NFW profile), 
$\rho_c(z) = 3H(z)^2/8\pi G$ is the critical density of the universe at the redshift $z$, $R_s$ is 
the scale radius, and $\delta_c$ is a characteristic dimensionless density; when expressed in 
terms of the concentration parameter $c$, $\delta_c$ takes the form
\begin{equation}
\delta_c = \frac{200}{3}\frac{c^3}{\ln(1+c) - c/(1+c)}. \label{eqn.conc}
\end{equation}
In NGC 1407 this phenomenon is also observed. The total mass profile of NGC 1407 in $>0.85$\re, 
however, can be well fitted with the NFW model. The obtained best-fit parameters are 
$r_s=3.4\pm0.4$\re~and $c=18.6\pm1.5$. 
The inferred virial radius $r_{200}$, where the average mass density is 200 times the current 
critical density of the universe, is $r_{200}=572\pm118$ kpc. At $r_{200}$ the inferred virial 
mass and mass-to-light ratio are  
$M_{200}=2.20\pm0.42\times 10^{13}$ \solarM~
and
$M_{\rm vir}/L_{\rm B}=311\pm60$ \solarMtosolarLB, respectively.

\subsection{Dark Matter}
By subtracting the contributions of the black hole (\S3.2), stars (\S5.2) and gas (\S5.1) from 
the total mass, we calculated the dark matter distribution in NGC 1407 and its associated group.
We plot the dark mass profile in Figure 5c in which a slope change at $\sim0.85$\re~can be 
clearly seen. Neither the NFW profile nor the generalized NFW profiles can give an accepted fit 
to this abnormal profile, similar to what we have found for the total mass profile. We also have 
tentatively tested the model consisting of two mass components, where each component takes the 
form of the NFW, the generalized NFW or the beta profile. None of these combinations can give 
an acceptable fit to the global dark matter distribution. For regions outside 0.85 \re, 
where the dark component begins to dominate the gravity, we find that the generalized NFW profile 
with $\zeta=0$ or $\zeta=1$ (=NFW) can shape the dark matter profile well, with 
$r_s=5.6$\re~for $\zeta=0$) or 9.3\re~for $\zeta=1$.
In Figure 5d we illustrate the dark matter fraction $f_{\rm dark}$ as a function of radius, 
along with the 90\% errors. We find that in the central region 
($^{<}_{\sim}0.85$\re~or 7.7 kpc)
the lower and upper limits of \totMtoLB~are about 5 and 10 \solarMtosolarLB, respectively,
inferring that the dark matter fraction is less than about 50\%. In the outer regions, however, 
the dark matter fraction increases fast outwards and begins to dominate the total gravitating 
mass at about 2\re, where the dark matter fraction is $80\pm5$\%.

A reliable description of the dark matter profile in early-type galaxies depends on to what extent 
errors and uncertainties on the stellar and gas mass are minimized. For NGC 1407, although the 
resulting stellar population mass-to-light ratio determined from absorption-line strengths appears 
to be a reasonable approximation to the dynamical mass-to-light ratio (see, e.g., Cappellari et al. 2006), 
and there is no apparent radio, optical or X-ray evidence against the assumption of the hydrostatic 
equilibrium, we would like to emphasize that the results still needs to be inspected further in the 
future by comparing with those derived from other independent methods

\section{Discussion}
\subsection{Errors and Uncertainties on Stellar Mass-to-Light Ratio}
We find that within 0.5\re~the calculated B-band mass-to-light ratio is nearly consistent with 
a constant distribution of 5.3\solarMtosolarLB~ (Thomas ${\rm [MgFe]_{T}}$) or 4.8\solarMtosolarLB~
(Cappellari ${\rm [MgFe]_{C}}$) at the 90\% confidence level (Table 3), if the Kroupa IMF is 
assumed. We also have crosschecked the results by adopting the diet-Salpeter IMF (Bell et al. 2003) 
and the Chabrier (2003) IMF. The results are essentially the same as those obtained with the Kroupa 
IMF, with a deviation of less than 10\%. If we convert the obtained B-band stellar mass-to-light 
ratios into their I-band counterparts, our results agree very well with the Cappellari et al.'s mean 
value for NGC 1407 (2.44\solarMtosolarLI). The calculated stellar mass-to-light ratios by using the 
Salpeter IMF are systematically larger by about a factor of 1.7. This should be ascribed to the fact 
that the Salpeter IMF overestimates the number of low-mass stars, as has been proposed by, e.g., 
Weiner et al. (2001) and Cappellari et al. (2006). No matter which IMF is used, the derived age 
is in excellent agreement with those in earlier works (Thomas et al. 2005; Howell 2005; 
Humphrey et al. 2006a), and is also consistent with the mean age of the early-type galaxies 
at $z=0$ that is obtained with the Spitzer data (Bregman et al. 2006).

As we found in \S5.2, the stellar population in NGC 1407 is $\alpha$-enriched, so we introduced
a combinative index MgFe as a reasonable tracer of the total metallicity of the stellar population.
To crosscheck our results on the stellar age and metallicity with an alternative approach, we 
also have interpolated the H$\beta$, Mgb, Fe5270, Fe5335 and Fe5015 index values given in the 
model table in Thomas et al. (2003), where the $\alpha$-products and iron are treated separately 
so that the impact of the $\alpha/{\rm Fe}$ bias is minimized. The derived ages are in good 
agreement with our PEGASE results, although the derived metallicities are marginally consistent 
with, or slightly higher than the PEGASE values due to the bluer red giant branches in the 
stellar evolutionary tracks that are used, as well as the higher adopted $\alpha/{\rm Fe}$ (0.3) 
comparing to the value (0) in PEGASE code. Within the frame of the simple stellar population 
model (Maraston 2005), these values infer quite a similar \starMtoLB~distribution to what we 
have obtained with the PEGASE code.

In addition to the possible systematic errors and uncertainties that are discussed above, there 
are a number of added error and uncertainty sources in the calculations of stellar mass-to-light 
ratios. In particular, by comparing the results obtained with the Kroupa IMF, the diet-Salpeter 
IMF and the Chabrier IMF, we estimate that the model uncertainties on the IMF profile may introduce 
a typical error of about 10\%. The size of this error is similar to that caused by the systematic 
errors or model uncertainties of the built-in stellar spectrum library of the Pegase code, e.g., 
the uncertainty in the stellar effective temperatures. Additional errors of about 5\% also come 
from the modeling of the star formation history and the evolution of galactic winds. 
Combining these errors with those occurred in the measurements of photometric data and line indices 
($5-10\%$), the resulting stellar mass-to-light ratios have a typical error of about $15-20\%$. 
More detailed correlative descriptions about the errors in the single stellar population models 
can be found in Trager et al. (2000) and Le Borgne et al. (2004), where consistent conclusions were 
given.

\subsection{Total Mass-to-Light Ratio and Dark Matter}
Like in other early-type galaxies (e.g., Lintott et al. 2005) and spiral galaxies (e.g., Kormendy and Bender 1999), 
we find that NGC 1407 has a baryon-dominated core. At 1\re, the mass-to-light ratio and dark 
matter fraction are estimated to be \totMtoLB~=7.7\solarMtosolarLB~and $f_{\rm dark}=43^{+14}_{-12}\%$, 
respectively, which is typical for elliptical galaxies (e.g., Cappellari et al. 2006; 
Fukazawa et al. 2006; Padmanabhan et al. 2004). In $>1$\re, \totMtoLB~and $f_{\rm dark}$ 
increases outwards quickly, which reaches 
\totMtoLB~=$22.2^{+7.8}_{-5.9}$\solarMtosolarLB~and 
$f_{\rm dark}=75^{+5}_{-5}\%$ at 2\re,
$46.2^{+20.3}_{-23.2}$\solarMtosolarLB~and $90^{+2.8}_{-2.0}\%$ at 5\re, 
and
$91.0^{+17.4}_{-56.3}$\solarMtosolarLB~and $95^{+1.2}_{-1.3}\%$ at 10\re.
This clearly implies the existence of a large amount of dark matter that dominating the gravitating 
mass. The measured \totMtoLB~of NGC 1407 is close to that found in NGC 1600 by Sivakoff et al. 
(2004; \totMtoLB~=$10\pm3$ and $24\pm6$\solarMtosolarLB~at about 1 and 4\re, respectively), who 
used the same X-ray technique as ours. Our \totMtoLB~is also close to, or significantly larger 
than those for NGC 3923 (\totMtoLB~is between $17-32$\solarMtosolarLB~at about 10\re) and  
NGC 720 (6.0, 12.3 and 18.7\solarMtosolarLB~at 1, 2 and 3\re, respectively), as were found by 
Buote and Canizares (1998) and Buote et al. (2002) who applied a geometric test technique to 
the \ROSAT~and \Chandra~images. Apparently, our \totMtoLB~of NGC 1407 is significantly larger 
than the controversial results of NGC 821 ($13-17$\solarMtosolarLB), NGC 3379($5-8$\solarMtosolarLB), 
NGC 4494 ($5-7$\solarMtosolarLB) and NGC 4697 ($\sim11$\solarMtosolarLB), which was measured 
in $4-6$\re~by Romanowsky et al. (2003) with the planetary nebula spectrograph.

At about 10\re~the inferred total mass are $6.43^{+1.23}_{-3.98}\times10^{12}$ \solarM, 
which are 1-6 times larger than the typical values for groups of galaxies (Humphrey et al. 2006b). 
At the virial radius $R_{200}$, the extrapolated gravitating mass and mass-to-light ratio are 
$2.20\pm0.42\times10^{13}$ \solarM~and $311\pm60$ \solarMtosolarLB, 
respectively. We find that these values are marginally consistent with those of Quintana et al. (1994) 
and are slightly larger than those of Humphrey et al. (2006b). Our results are lower by a factor 
of $2-3$ than those of Trentham et al. (2006) and Gould (1993) in which NGC 1400 was included 
in the kinematic analysis. This casts some doubts on the identity of NGC 1400 as a virialized 
member in the group. Since NGC 1400 is likely to be at the distance of the NGC 1407 group, as 
was suggested by Perrett et al. (1996) who compared the shapes of GC luminosity functions of 
NGC 1400 and NGC 1407, the observed unusually large peculiar velocity may imply that NGC 1400 
is intruding into and merging with the group. By comparing the mass-to-light ratio of the NGC 1407 
group with those listed in, e.g., Sanderson \& Ponman (2003), we find that, similar to the NGC 1600 
group, the NGC 1407 group is an extremely dark system even comparable to many clusters of galaxies.

We find that the total mass distribution in the NGC 1407 group shows a slop change, or a 
flattened feature at about $^{<}_{\sim}0.85$\re~(7.7 kpc). Outside this radius the dark 
matter fraction increases rapidly outwards. This is in consistence with the result in the 
sample work of Humphrey et al. (2006b). This flattened feature may be caused by the change 
of the relative contribution of the gas mass as compared with the stellar mass. However, 
we notice that a similar feature is also seen on the mass profile of dark matter, to derive 
which the contributions of both bright and dwarf members have been carefully taken into account. 
To examine if the appearance of the flattened feature is model-dependent, we have repeated 
our calculations in \S5.1, \S5.3 and \S5.4 by starting from the deprojected X-ray surface 
brightness profile (e.g., Chatzikos et al. 2006), to derive which the background given in 
\S4.1 is used. We find that within the errors the results agree nicely with what we presented 
above, indicating that the flattened, or the shoulder-like feature is intrinsic. 
We speculate that the slope changes on both the \totM~and \totMtoLB~profiles are likely 
to be a mark of the boundary between the galaxy and group halos, since in $1-2$\re~we also 
observe a transition of the gas temperature from the galactic level ($\simeq0.7$ keV) to the 
group level ($>1$ keV). Similar phenomenon and its analogs have been observed in clusters 
of galaxies (Ikebe 1995); Xu et al. (1998); Ettori et al. 2002; also Makishima et al. 2001 
and references therein), as well as in other groups (NGC 4472 and NGC 4649, Brighenti \& Mathews 1997; 
NGC 507, Paolillo et al. 2003; NGC 1600, Sivakoff et al. 2004; IC 1459 and NGC 1399, Fukazawa et al 2006; 
NGC 4125 and NGC 4261, Humphrey et al. 2006b). We find that in more than half of these cases 
the slope change on mass profile appears at $7\sim10$ kpc, meanwhile in roughly half of the 
cases an outward temperature increase is observed. We also notice that the slope change 
is always seen in the cluster/group-dominating galaxies, or galaxies dominating a sub-cluster. 
In the isolated or nearly isolated elliptical galaxies, such as NGC 1404 (Paolillo et al. 2002),
NGC 4555 (O'Sullivan \& Ponman 2004), NGC 4636 (Brighenti \& Mathews 1997) and NGC 6482 (Humphrey et al. 2006b), 
such a feature is not clearly detected. Because NGC 6482 is believed to dominate a fossil group,
the observed slope change on mass profile may finally vanish when the halo collapse enters the
final stage.

The mass distribution of NGC 1407 and its group in $>1$\re, where the mass is dominated 
by dark matter, can be well fitted by a single NFW profile. The best-fit parameter are 
$r_s=1.3$\re~($\zeta=0$) or 3.5 \re~($\zeta=1$). The derived concentration 
parameter c ($18.6\pm1.5$) is within the range allowed for groups, but is beyond the 68\% 
scatter for a halo at $z=0$ at the given $M_{\rm vir}$ (Gnedin and Ostriker 2001; Jing 2000). 
Despite of the relatively large errors we find that the inferred total mass and dark matter 
distributions within $0.85$\re~are cuspy and can be approximated by power-law profiles 
with indices of $\sim2$, which are marginally consistent with the generalized NFW profiles
with $\zeta=2$.

\subsection{Stellar Metallicity Gradient}
Like what was found in some other giant elliptical galaxies (Arimoto et al. 1997 and references 
therein), the calculated stellar metallicity in NGC 1407 is not spatially uniform. This is 
mostly determined by the observed spatial variations of the metal line indices, and is almost 
independent of the model assumptions on the IMF profile and so on. The inward metallicity 
enhancement has been predicted by the monolithic collapse models (e.g., Eggen et al. 1962). 
Within the frame of such models, more metal-rich stars were born in the inward-dissipating 
gas that had been enriched on their way towards the galactic center, meanwhile the evolving 
stellar winds tend to expel more metals out of the outer regions where the gravitational 
potential is shallow. Numerical simulations show that the stellar metallicity gradients are 
also adjusted by mergers. Although the mergers tend to smear out any existing gradients, 
they also may supply a large amount of gas and trigger star formation in the inner regions. 
So the resulting metallicity gradients are not necessary to be as steep as is predicted by 
the monolithic collapse models (e.g., Sanchez-Blazquez et al. 2006 and references therein), 
as is observed in NGC 1407. 
If the stellar metallicity gradient at $\sim0.5$\re~ is indeed adjusted by gas inflows, by 
calculating the time that the gas therein needed to radiatively cool down to reach the current 
temperature ($\simeq0.7$ keV), which is about 4 Gyr, we tentatively estimate that such a gas 
inflow event should occurred at least 4 Gyr ago. From then on the gas therein has sufficient 
time to reestablish an equilibrium state.

\section{Summary}\label{sec:summary}

By incorporating the \Chandra~and \ROSAT~X-ray spectroscopic data with the optical line-strength
indices and multi-color photometric data, we studied the mass distributions in the bright E0 galaxy
NGC 1407 and its associated group. We find that the gas is single-phase with a temperature of
$\simeq0.7$ keV within 1\re~(1\re~= 9.0 kpc), which quickly increases to $>1$ keV in $1-2$\re.
We reveal that the X-ray surface brightness profile shows a central excess in the innermost region,
which coincides with a flattened feature at about $^{<}_{\sim}1$\re~(7.7 kpc) on the total mass
and dark matter profiles. This feature is speculated to be a mark of the boundary between the galaxy
and group halos, as has been seen in some other cluster/group-dominating galaxies. We find that the
total mass and dark matter distributions within $0.85$\re~(7.7 kpc) are cuspy and can be approximated
by power-law profiles with indices of $\sim2$, which are marginally consistent with the generalized
NFW profiles. The mass in outer regions can be well fitted by a single NFW profile, and the derived
concentration parameter c ($18.6\pm1.5$) is larger than the 68\% upper limit for a halo at $z=0$ at
the given $M_{\rm vir}$. We conclude that although the NGC 1407 group has a baryon-dominated core,
it is an extremely dark system even comparable to many clusters of galaxies. At the virial radius
$r_{200}=572\pm118$ kpc, the calculated mass and mass-to-light ratio are 
$M_{200}=2.20\pm0.42 \times 10^{13}$ \solarM~and $M_{\rm vir}/L_{\rm B}=311\pm60$ \solarMtosolarLB,
respectively. We also argue that the unusually large peculiar velocity of NGC 1400 may reflect the
fact that the galaxy may have not be virialized in the group's gravitational potential well.

\acknowledgments

We thank Roberto Rampazzo and Annibalii for kindly providing us with the optical line-strength indices 
that are used in this paper. We thank Damien Le Borgne, Claudia Maraston and Scott C. Trager for 
their helpful suggestions and comments. This work was supported by the National Science Foundation 
of China (Grant No. 10273009, 10233040 and 10503008), Shanghai Key Projects in Basic Research No. 04JC14079, 
and the Ministry of Science and Technology of China, under Grant No. NKBRSF G19990754.


\clearpage

\begin{deluxetable}{lcccccccr}
\tabletypesize{\scriptsize}
\tablecaption{\Chandra~and \ROSAT~Spectral Analyses}
\tablewidth{0pt}
\tablehead{
& & 
\multicolumn{2}{c}{Non-thermal Component} & &
\multicolumn{3}{c}{Thermal Component} &\\
\cline{3-4} 
\cline{6-8}
\colhead{Source} &
\colhead{Model\tablenotemark{a} \,} &
\colhead{$\Gamma$ or $kT_{\rm in}$} & 
\colhead{$F_{\rm X}$\tablenotemark{b} \,} & &
\colhead{$kT$} & 
\colhead{$Z$} &
\colhead{$F_{\rm X}$\tablenotemark{b} \,} &\\
& &
\colhead{(keV)} & & &
\colhead{(keV)} & 
\colhead{(solar)} & &
\colhead{$\chi^2$/d.o.f}
}

\startdata
\multicolumn{9}{c}{(1) Central Source}\\
\hline
 Src 1\tablenotemark{c} \, &PL   & $2.29^{+0.24}_{-0.23}$ & $0.47^{+0.05}_{-0.06}$ & &\nodata &\nodata &\nodata & 10.2/11\\
                         &DBB  & $0.49^{+0.15}_{-0.10}$ & $0.30^{+0.15}_{-0.14}$ & &\nodata &\nodata &\nodata & 27.8/11\\

\hline
\multicolumn{9}{c}{(2) \Chandra~Projected Analysis (0.7--7 keV)\tablenotemark{d} \,}\\
\hline

1 &PL+APEC   &1.59/fixed &$1.26\pm0.07$ & &$0.68\pm0.02$ &$0.52^{+0.11}_{-0.15}$ &$3.31\pm0.92$ &71.0/48\\
2 &PL+APEC   &1.59/fixed &$0.70^{+0.22}_{-0.09}$ & &$0.73\pm0.02$ &$0.55^{+0.18}_{-0.14}$ &$3.04\pm0.86$ &82.5/48\\
3 &PL+APEC   &1.59/fixed &$1.31^{+0.17}_{-0.15}$ & &$0.91\pm0.03$ &$0.73^{+0.43}_{-0.18}$ &$2.05\pm0.69$ &62.2/48\\      
4 &PL+APEC   &1.59/fixed &$1.97^{+0.47}_{-0.31}$ & &$1.21^{+0.09}_{-0.12}$ &$0.38^{+0.27}_{-0.14}$ &$2.24\pm0.69$ &52.7/48\\ 

\hline
\multicolumn{9}{c}{(3) \Chandra~Deprojected Analysis (0.7--7 keV)\tablenotemark{d} \,}\\
\hline
1 &PL+APEC   &1.59/fixed &\nodata & &$0.66\pm0.02$ &$0.60^{+0.41}_{-0.20}$ &\nodata &617/634\\
2 &PL+APEC   &1.59/fixed &\nodata & &$0.68\pm0.02$ &$0.63^{+0.46}_{-0.20}$ &\nodata &\nodata\\
3 &PL+APEC   &1.59/fixed &\nodata & &$0.85^{+0.04}_{-0.02}$ &$1.34\pm0.57$ &\nodata &\nodata\\  
4 &PL+APEC   &1.59/fixed &\nodata & &$1.25\pm0.09$ &$0.52^{+0.31}_{-0.25}$ &\nodata &\nodata\\ 

\hline
\multicolumn{9}{c}{(2) \ROSAT~Projected Analysis (0.2--2 keV)\tablenotemark{e} \,}\\
\hline

1 &PL+APEC   &1.59/fixed &$0.76\pm0.52$ & &$0.66^{+0.09}_{-0.12}$ &$0.73^{+0.13}_{-0.11}$ &$1.55\pm0.24$ &82.9/69\\
2 &PL+APEC   &1.59/fixed &$0.42\pm0.32$ & &$0.67^{+0.06}_{-0.07}$ &$0.98^{+0.13}_{-0.11}$ &$3.53\pm0.25$ &63.0/69\\
3 &PL+APEC   &1.59/fixed &$0.78\pm0.08$ & &$0.64^{+0.09}_{-0.10}$ &$1.44\pm0.52$ &$1.69\pm0.55$ &83.0/69\\      
4 &PL+APEC   &1.59/fixed &$1.24\pm0.37$ & &$1.03^{+0.30}_{-0.20}$ &$0.69^{+0.46}_{-0.18}$ &$2.23\pm0.27$ &70.4/69\\ 
5 &APEC      &\nodata &\nodata                          & &$1.37^{+0.55}_{-0.32}$ &$0.80^{+0.67}_{-0.35}$ &$2.32\pm0.31$ &51.2/69\\
6 &APEC      &\nodata &\nodata                          & &$1.50\pm0.48$ &$0.21\pm0.07$ &$2.69\pm0.46$ &185.0/209\\
7 &APEC      &\nodata &\nodata                          & &$1.09^{+0.42}_{-0.19}$ &0.21/tied              &$4.66\pm0.68$ &\nodata\\      
8 &APEC      &\nodata &\nodata                          & &$1.09\pm0.35$ &0.21/tied              &$2.63\pm0.75$ &\nodata\\

\hline
\multicolumn{9}{c}{(3) \ROSAT~Deprojected Analysis (0.2--2 keV)\tablenotemark{e} \,}\\
\hline
1 &PL+APEC   &1.59/fixed &\nodata & &$0.73^{+0.27}_{-0.19}$ &$0.59\pm0.17$ &\nodata &531/576\\
2 &PL+APEC   &1.59/fixed &\nodata & &$0.69\pm0.08$ &$0.78\pm0.08$ &\nodata &\nodata\\
3 &PL+APEC   &1.59/fixed &\nodata & &$0.63\pm0.10$ &$1.33^{+0.14}_{-0.18}$ &\nodata &\nodata\\  
4 &PL+APEC   &1.59/fixed &\nodata & &$1.12^{+0.44}_{-0.32}$ &$0.60^{+0.59}_{-0.22}$ &\nodata &\nodata\\ 
5 &APEC      &\nodata    &\nodata & &$1.37\pm0.35$ &$0.49^{+0.64}_{-0.25}$ &\nodata &\nodata\\
6 &APEC      &\nodata    &\nodata & & 1.37/tied             &$0.21^{+0.06}_{-0.07}$ &\nodata &\nodata\\
7 &APEC      &\nodata    &\nodata & &$1.22\pm0.35$ & 0.21/tied             &\nodata &\nodata\\  
8 &APEC      &\nodata    &\nodata & &$1.08\pm0.30$ & 0.21/tied             &\nodata &\nodata\\

\enddata

\tablenotetext{a}{Models used in the spectral fittings are 
PL = the power-law model,
DBB = the multiple color disk model (diskbb) and
APEC = the thermal plasma model. 
In all cases the column density are fixed to the Galactic value $5.42\times10^{20}$ cm$^{-2}$ 
(Dickey \& Lockman, 1990), since allowing it to vary does not improve the fit.}

\tablenotetext{b}{0.7--7 keV fluxes for the \Chandra~observation or 0.2--2 keV fluxes for \ROSAT~observation
in $10^{-13}$ ergs cm$^{-2}$ s$^{-1}$ (90\% errors), which have been corrected for absorption.}
\tablenotetext{c}{The central source.}
\tablenotetext{d}{Annuli used to extract the \Chandra~ACIS S3 spectra: 
$0-0.25$\re,
$0.25-0.53$\re,
$0.53-0.91$\re~and
$0.91-2.1$\re.}
\tablenotetext{e}{Annuli used to extract the \ROSAT~PSPC spectra: 
$0-0.25$\re,
$0.25-0.53$\re,
$0.53-0.91$\re,
$0.91-2.1$\re,
$2.1-3.4$\re,
$3.4-5.1$\re,
$5.1-7.7$\re~and
$7.7-10.3$\re.
}

\end{deluxetable}

\begin{deluxetable}{cccccccc}
\tabletypesize{\scriptsize}
\tablecaption{Line-strength Indices\tablenotemark{a} \,and Multi-color Photometric Data\tablenotemark{b} \,}
\tablewidth{0pt}
\tablehead{
\colhead{Radius} &
\colhead{H$\beta$} &
\colhead{Mgb} & 
\colhead{Fe5015} & 
\colhead{Fe5270} &
\colhead{Fe5335} & 
\colhead{B-R} & 
\colhead{B-K}\\
\colhead{(\re)} &
\colhead{(\r{A})} &
\colhead{(\r{A})} & 
\colhead{(\r{A})} & 
\colhead{(\r{A})} &
\colhead{(\r{A})} & 
\colhead{(mag)} & 
\colhead{(mag)}
}

\startdata
0-1/16   & $1.48\pm0.10$ & $5.29\pm0.08$ & $6.37\pm0.21$ & $3.25\pm0.11$ & $2.75\pm0.14$ & $1.63\pm0.14$ & $4.59\pm0.14$\\
1/16-1/8 & $1.67\pm0.10$ & $5.28\pm0.09$ & $6.11\pm0.21$ & $3.08\pm0.12$ & $2.44\pm0.14$ & $1.61\pm0.14$ & $4.44\pm0.14$\\
1/8-1/4  & $2.05\pm0.10$ & $5.24\pm0.09$ & $6.11\pm0.21$ & $3.19\pm0.12$ & $2.72\pm0.14$ & $1.61\pm0.14$ & $4.34\pm0.14$\\
1/4-1/2  & $1.81\pm0.10$ & $5.05\pm0.09$ & $5.68\pm0.21$ & $2.30\pm0.12$ & $1.88\pm0.14$ & $1.60\pm0.14$ & $4.30\pm0.14$\\
\enddata
\tablenotetext{a}{The line-strength indices transformed into the Lick-IDS system (Rampazzo et al. 2005).}
\tablenotetext{b}{B- and R-band photometric data are obtained from de Carvalho et al. (1991), while K-band photometric data are from Jarrett et al. (2003).}

\end{deluxetable}


\begin{deluxetable}{cccccccc}
\tabletypesize{\scriptsize}
\tablecaption{Stellar Ages, Metallicities and Mass-to-Light Ratios}
\tablewidth{0pt}
\tablehead{
\colhead{Radius} &
\colhead{\starMtoLB} &
\colhead{Z\tablenotemark{b} \,} & 
\colhead{age} & &
\colhead{\starMtoLB} &
\colhead{Z\tablenotemark{b} \,} & 
\colhead{age} \\
\colhead{(\re\tablenotemark{a} \,)} &
\colhead{(\solarMtosolarLB)} &
\colhead{($Z_{\odot}$)} & 
\colhead{(Gyr)} & &
\colhead{(\solarMtosolarLB)} &
\colhead{($Z_{\odot}$)} & 
\colhead{(Gyr)} 
}

\startdata
\multicolumn{8}{c}{PEGASE code}\\
\hline
& 
\multicolumn{3}{c}{Salpeter IMF+${\rm [MgFe]_{T}}$} & & 
\multicolumn{3}{c}{Kroupa IMF+${\rm [MgFe]_{T}}$} \\
\cline{2-4} 
\cline{6-8}
0-1/16   &$9.8^{+1.1}_{-1.3}$ & $1.7\pm0.4$ &$9.7^{+0.8}_{-0.9}$  && $5.8^{+0.8}_{-0.7}$ & $1.5^{+0.4}_{-0.5}$ & $10.4^{+1.1}_{-0.8}$\\
1/16-1/8 &$9.2^{+1.6}_{-1.4}$ & $1.3^{+0.7}_{-0.3}$ &\nodata                  && $5.5^{+1.1}_{-0.7}$ & $1.1^{+1.0}_{-0.6}$ & \nodata \\
1/8-1/4  &$9.4^{+1.2}_{-1.1}$ & $2.6\pm0.2$ &\nodata                  && $5.6^{+1.0}_{-0.7}$ & $2.5^{+0.2}_{-0.3}$ & \nodata \\
1/4-1/2  &$7.4\pm1.2$ & $0.5^{+0.3}_{-0.2}$ &\nodata                  && $4.3^{+0.9}_{-0.5}$ & $0.3^{+0.2}_{-0.1}$ & \nodata \\
\hline
& 
\multicolumn{3}{c}{Salpeter IMF+${\rm [MgFe]_{C}}$} & & 
\multicolumn{3}{c}{Kroupa IMF+${\rm [MgFe]_{C}}$} \\
\cline{2-4} 
\cline{6-8}

0-1/16   &$8.9^{+0.9}_{-1.0}$ & $1.8\pm0.3$ &$8.5^{+1.0}_{-0.7}$  && $5.0^{+0.7}_{-0.6}$ & $1.6\pm0.3$ & $8.8^{+1.0}_{-1.3}$\\
1/16-1/8 &$8.7^{+1.1}_{-1.3}$ & $1.6^{+0.6}_{-0.4}$ &\nodata                  && $5.0\pm0.7$ & $1.5\pm0.5$ & \nodata \\
1/8-1/4  &$8.3^{+1.2}_{-0.6}$ & $2.6^{+0.3}_{-0.5}$ &\nodata                  && $4.8^{+0.9}_{-0.4}$ & $2.4^{+0.2}_{-0.3}$ & \nodata \\
1/4-1/2  &$7.8^{+1.3}_{-1.1}$ & $1.0\pm0.3$ &\nodata                  && $4.6^{+0.9}_{-0.8}$ & $0.9^{+0.5}_{-0.3}$ & \nodata \\
\hline
\multicolumn{8}{c}{Thomas and Maraston models\tablenotemark{c} \,} \\
\hline
0-1/16   &$10.0^{+1.2}_{-1.1}$ & $2.5\pm0.2$ & $9.2\pm0.7$ && $6.3^{+0.8}_{-0.6}$ & $2.5\pm0.2$ & $9.2\pm0.7$\\
1/16-1/8 &$9.8^{+1.2}_{-1.1}$ & $2.3^{+0.2}_{-0.3}$ &\nodata                  && $6.2^{+0.8}_{-0.7}$ & $2.3^{+0.2}_{-0.3}$ & \nodata \\
1/8-1/4  &$9.8^{+1.2}_{-1.1}$ & $2.3^{+0.2}_{-0.3}$ &\nodata                  && $6.2^{+0.8}_{-0.7}$ & $2.3^{+0.2}_{-0.3}$ & \nodata \\
1/4-1/2  &$8.8^{+1.2}_{-1.1}$ & $1.7\pm0.2$ &\nodata                  && $5.6\pm0.7$ & $1.7\pm0.2$ & \nodata \\

\enddata

\tablenotetext{a}{Effective radius (1\re~=$1.17^{\prime}$).}
\tablenotetext{b}{Calculated stellar total abundance.}
\tablenotetext{c}{Stellar ages and abundances are calculated by using the Thomas model (Thomas et al. 2003). \starMtoLB~is calculated based on the Maraston model (Maraston 2005).}
\end{deluxetable}

\clearpage

\begin{figure}
\epsscale{1.0}
\begin{center}
\includegraphics[width=12.0cm,angle=0]{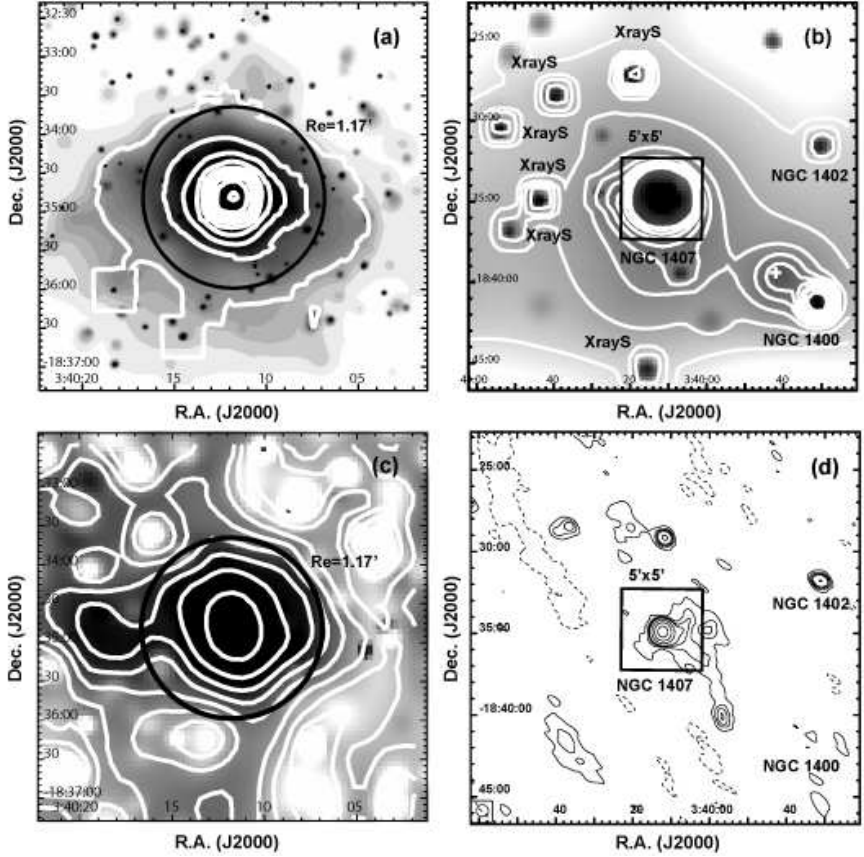}
\end{center}
\figcaption{
(a): \Chandra~ACIS S3 image in 0.3--10 keV in logarithmic scale with the intensity contours 
overlapped. The image has been corrected for exposure but not for background, and has been 
smoothed by using a minimum signal-to-noise ratio of $3\sigma$ and a maximum signal-to-noise 
ratio of $5\sigma$ per beam. The circle on the image represents the region of 1\re. 
All detected point sources ($3\sigma$) are removed.
(b): \ROSAT~PSPC image in 0.2--2 keV in logarithmic scale with the intensity contours 
overlapped. The image has been corrected for exposure but not for background, and has been 
smoothed in the same way as in Figure 1a. The \Chandra~observation field is marked with a box. 
The cross indicates the weak, extented structure aligned in the NGC 1407-NGC 1400 direction.
(c): Hardness ratio distribution that has been smoothed in the same way as in Figure 1a. The 
hardness ratio is defined as the ratio of exposure- and background-corrected counts in 0.3--2 keV 
to those in 2-10 keV, which ranges from -0.94 to 0.54 in this image.
(d): 1.43 GHz radio map obtained from the National Radio Astronomical Observatory (NRAO) archive, 
with peak flux density of $50.6~{\rm mJy/b}$ and contour levels of $(-1,1,2,3,4,8,16,32)~{\rm mJy/b}$.
\label{fig1}}
\end{figure}

\begin{figure}
\epsscale{1.0}
\begin{center}
\includegraphics[width=6.0cm,angle=270]{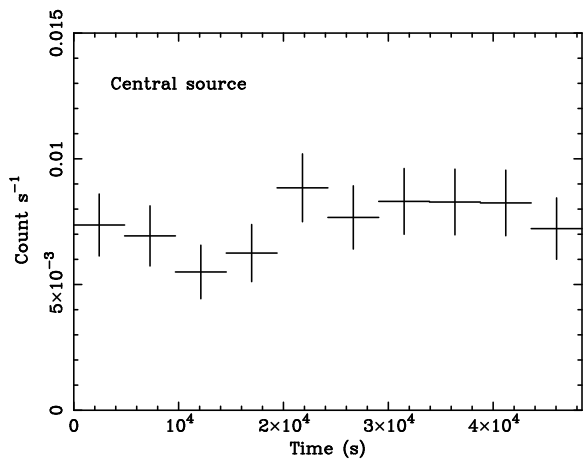}
\end{center}
\figcaption{Background-corrected lightcurve of the central point source. 
\label{fig2}}
\end{figure}

\begin{figure}
\epsscale{1.0}
\begin{center}
\includegraphics[width=8.0cm,angle=270]{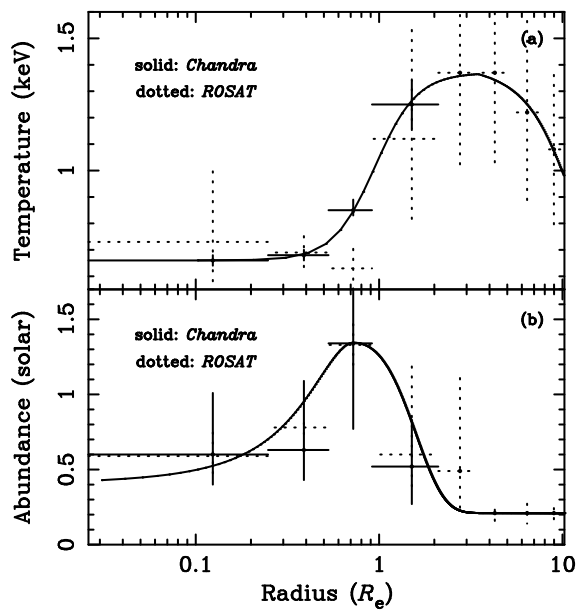}
\end{center}
\figcaption{Deprojected gas temperature distribution (90\% errors) and abundance distribution 
(68\% errors) obtained with \Chandra~ACIS S3 (solid crosses) and \ROSAT~PSPC (dotted crosses). 
Solid lines are the best-fit models. 
\label{fig3}}
\end{figure}

\begin{figure}
\epsscale{1.0}
\begin{center}
\includegraphics[width=12.0cm,angle=270]{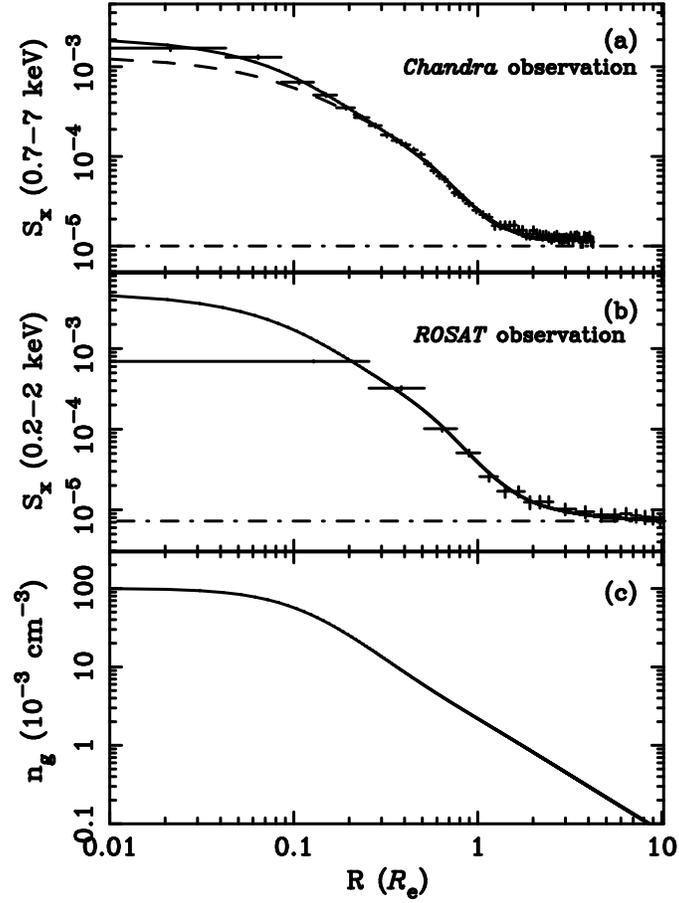}
\end{center}
\figcaption{
(a): The \Chandra~0.7--7 keV surface brightness profile (solid crosses) of the gas emission, which
has been corrected for exposure but not for background. The best-fit beta model and two-beta model are shown as 
a dashed line and a solid line, respectively.  
(b): The \ROSAT~0.2--2 keV surface brightness profile (solid crosses) of the gas emission, which
has been corrected for exposure but not for background. The best-fit two-beta model is shown 
as a solid line.  
(c): Calculated gas density distribution. 
\label{fig4}}
\end{figure}


\begin{figure}
\epsscale{1.0}
\begin{center}
\includegraphics[width=8.0cm,angle=270]{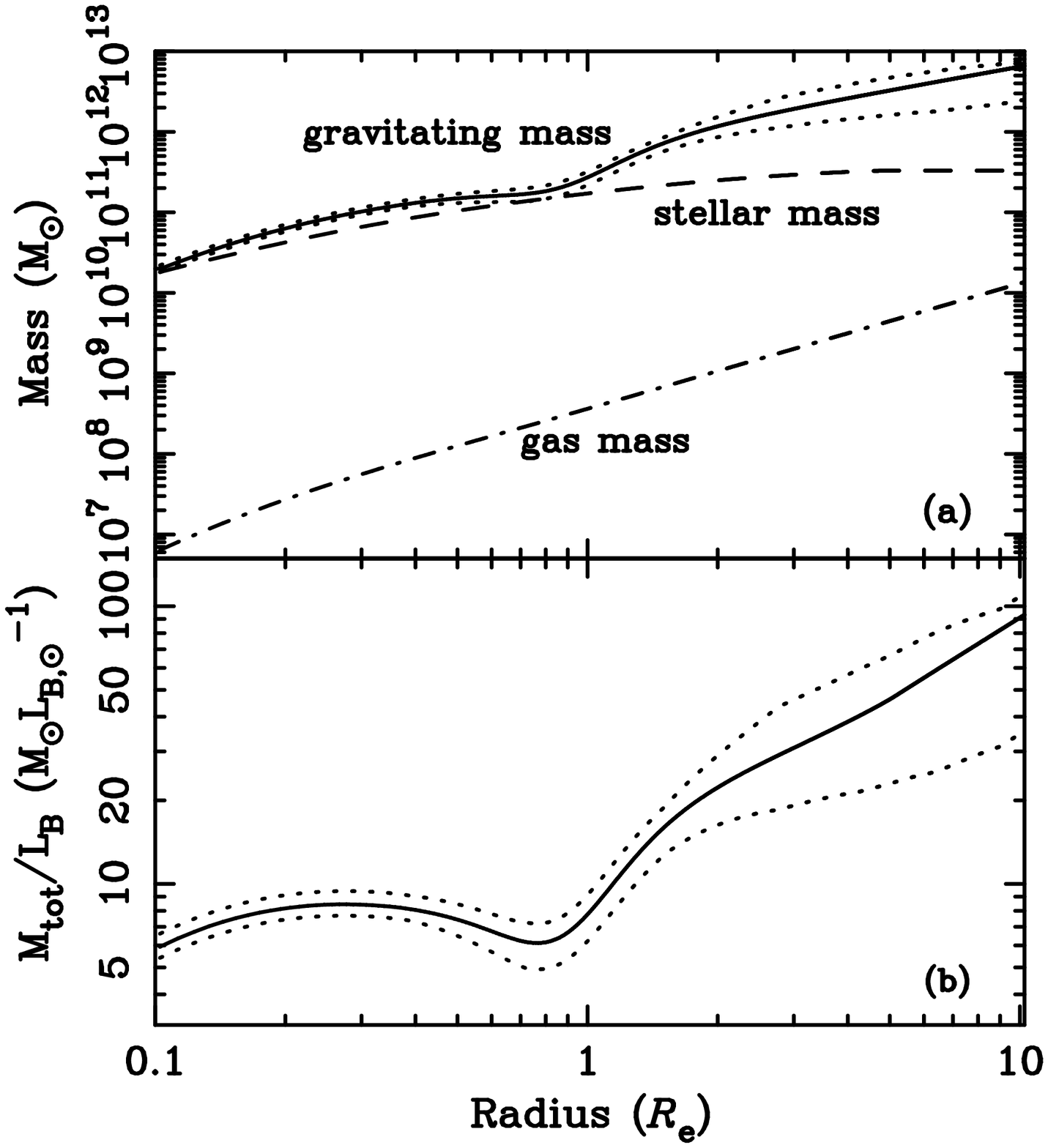}
\includegraphics[width=8.0cm,angle=270]{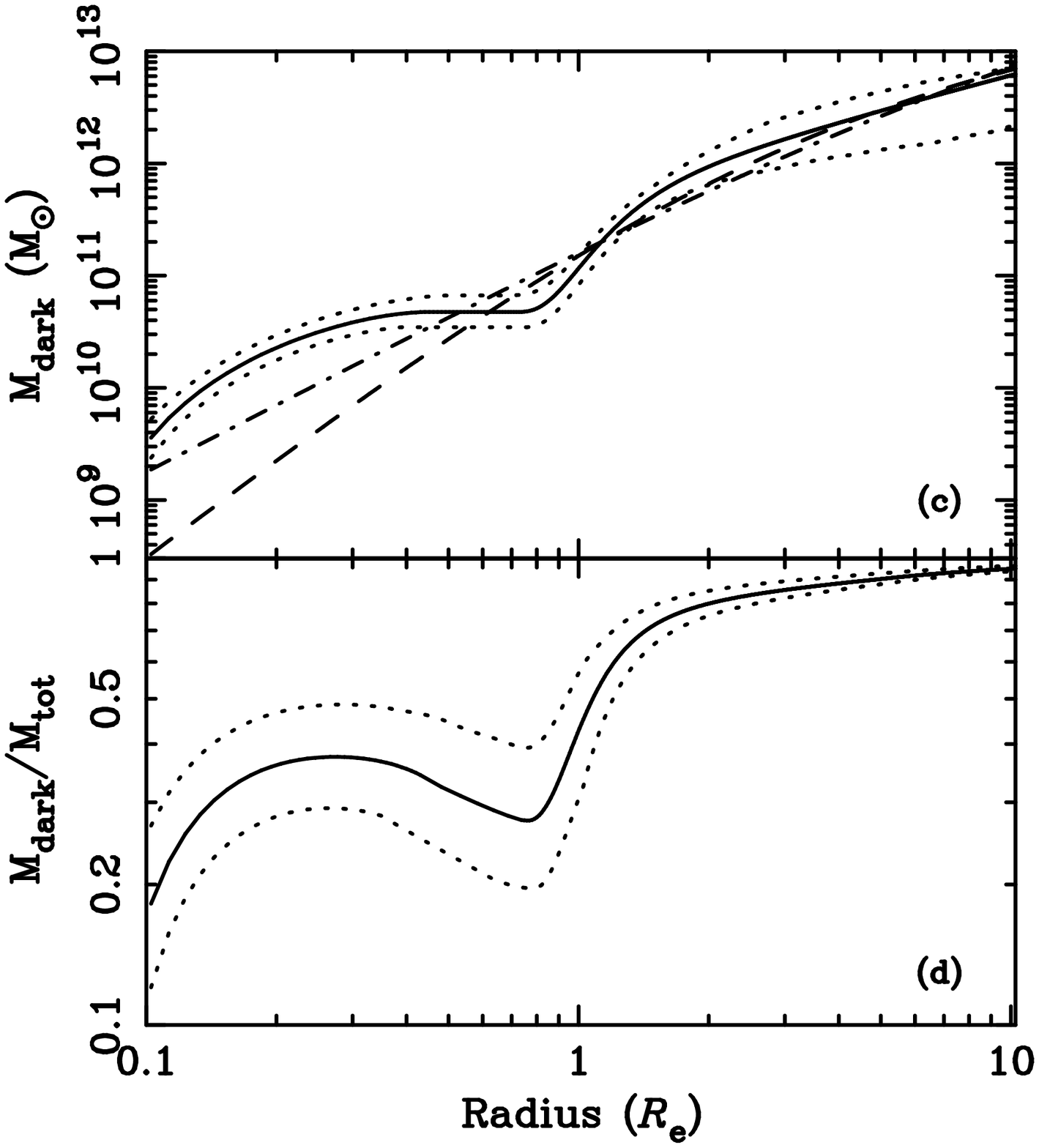}
\end{center}
\figcaption{
(a): Calculated total gravitating mass profile (solid line) and the 90\% errors (dotted lines), 
the gas mass profile (dash-dotted line) and the stellar mass profile (dashed line).
(b): Radial mass-to-light ratio distribution and the 90\% errors (dotted lines). 
(c): Radial dark matter distribution and the best-fit generalized NFW models for the $>0.85$\re~regions. 
The dashed line is for $\zeta=0$ and the dash-dotted line is for $\zeta=1$. Note that the models give a poor
fit to the mass profile of the innermost region.
(d): Darkmatter fraction (solid line) and the 90\% errors (dotted lines).
\label{fig5}}
\end{figure}

\end{document}